\documentclass[preprint,12pt]{elsarticle}

 \usepackage{epsfig}

 \usepackage{amssymb}

\journal{ }

\begin{document}
\begin{frontmatter}



\title{Sensitivity of the correlation between the depth of shower maximum and the muon shower size to the
cosmic ray composition}

\author[label1,label2,label3]{Patrick Younk\corref{cor1}}
\ead{pwyounk@gmail.com}
\author[label1]{Markus Risse}

\address[label1]{University of Siegen, D-57068 Siegen, Germany\\}
\address[label2]{Colorado State University, Fort Collins, Colorado 80523, USA}
\address[label3]{Los Alamos National Laboratory, Los Alamos, New Mexico 87545, USA}

\cortext[cor1]{Corresponding author at: Los Alamos National Laboratory, Los Alamos, New Mexico 87545, USA, phone: +1 505 606 0594, fax: +1 505 665 4121}

\begin{abstract}
The composition of ultra-high energy cosmic rays is an important issue in
astroparticle physics research, and 
additional experimental results are required for
further progress.
Here we investigate what can be learned from the statistical
correlation factor $r$ between the depth of shower maximum and the muon shower size,
when these observables are measured simultaneously for a set of air showers.
The correlation factor $r$ contains the lowest-order moment of a two-dimensional
distribution taking both observables into account, and it is
independent of systematic uncertainties of the absolute scales
of the two observables.
We find that, assuming realistic measurement uncertainties, the value of $r$ can provide
a measure of the spread of masses in the primary beam.
Particularly, one can differentiate between a well-mixed composition
(i.e., a beam that contains large fractions of both light and heavy primaries)
and a relatively pure composition (i.e., a beam that contains
species all of a similar mass).
The number of events required for a statistically significant
differentiation is $\sim200$.
This differentiation, though diluted,
is maintained to a significant extent in the presence of uncertainties in 
the phenomenology of high energy hadronic interactions.
Testing whether the beam is pure or well-mixed is 
well motivated by recent measurements of the depth of shower maximum.
\end{abstract}

\begin{keyword}
%
%
Ultra high energy cosmic rays \sep Extensive air showers \sep Cosmic ray composition
%
%
%
\end{keyword}
\end{frontmatter}


\section{Introduction}
The composition of ultra-high energy (UHE) cosmic rays is a key question in
astroparticle physics research.
There is relatively little guidance from theory about what particles to expect.
Typically, particle masses ranging from protons to heavy nuclei, such as iron,
are regarded.
Experimental input is needed to clarify the situation.

Measurements of the depth of shower maximum $X_{\max}$ from the Pierre Auger
Observatory~\cite{XmaxPaper} show
a statistically significant flattening of the elongation rate near $2 \times 10^{18}$~eV.
In addition,
the natural shower-to-shower fluctuations of $X_{\max}$ 
are reported to decrease from approximately 
$60~\mbox{g}/\mbox{cm}^2$ at $2 \times 10^{18}$~eV
to $30~\mbox{g}/\mbox{cm}^2$ at $4 \times 10^{19}$~eV.
These observations suggest an increase of the average nuclear mass of
UHE cosmic rays with energy.
In particular, the shower-to-shower fluctuations observed above $10^{19}$~eV 
indicate a nuclear composition (i.e., some mixture of isotopes with
mass number $A \geq 4$) with a relatively small proton component.

However, the quantitative interpretation of the data in terms of 
cosmic ray composition requires a comparison to shower simulations.
Because the relevant energies have not been explored in laboratory experiments,
the use of interaction parameters extrapolated from lower energy
experiments is required.
For instance, a comparison
of recent LHC data to predictions of different hadronic interaction
models showed reasonable overall agreement, but each model had
its shortcomings \cite{Engel}. This model uncertainty introduces
an uncertainty in the interpretation
of the $X_{\max}$ measurements.


There are also other results that are less suggestive
of a nuclear composition at the highest energies.
The HiRes Collaboration has reported results \cite{HiResElong} that are compatible with a pure
proton beam.
In addition,
the correlation of the arrival directions of the highest energy
cosmic rays with nearby extragalactic matter observed by the Auger Observatory~\cite{Abraham2007} is
not unexpected if the particles are protons, for which magnetic deflections
of a few degrees are plausible, yet it is more difficult to explain if the
particles are heavy nuclei. 

Thus, the current situation is unclear.
Additional independent experimental
results related to composition appear mandatory for
further progress.



The muon shower size $N_{\mu}$ is known to be well correlated to the mass
of the primary cosmic ray, 
and a measurement of $N_{\mu}$ would generally be independent
of the measurement of $X_{\max}$.
In this way, it is natural to pursue the use of $N_{\mu}$ in composition studies.
However, because of significant model uncertainties,
obtaining robust composition information from $N_{\mu}$ is not straightforward.
To briefly demonstrate the problem, 
consider that the difference between the EPOS \cite{EPOS1, EPOS2} and Sibyll 2.1 \cite{Sibyll}
hadronic interaction models
with regard to the average muon shower size $\left< N_{\mu} \right>$
for $10^{19}$~eV proton showers
is $\sim40\%$.
For comparison, the expected difference between proton 
and iron showers with regard to $\left< N_{\mu} \right>$ is also $\sim40\%$.
Thus, at present, it is quite difficult to make a reliable inference from an observed
value of $\left< N_{\mu} \right>$ to the average nuclear mass.

This does not mean that all useful composition information
contained in the $N_{\mu}$ observable is muddled by model
uncertainties, but care must be taken as to what inferences to draw.
There may be other useful inferences about the primary composition
besides the average nuclear mass.

Here we investigate what can be learned from the {\it statistical
correlation} between the two observables $X_{\max}$ and $N_{\mu}$
assuming a set of events where both quantities are measured simultaneously.
Our definition of correlation (precisely defined in Section 2) contains the lowest-order moment of a two-dimensional
distribution taking both observables into account, and it is
independent of systematic uncertainties of the absolute scales
of the two observables.
To our knowledge, such a study has not been described in detail in the literature.

We find that the correlation between $X_{\max}$ and $N_{\mu}$ can provide
a measure of the spread of masses in the primary beam.
Particularly, one can differentiate between a well-mixed composition
(i.e., a beam with large fractions of both light and heavy primaries)
and a relatively pure composition (i.e., a beam that contains
species all of a similar mass). 
This differentiation remains significant and meaningful, even though diluted,
in the presence of model uncertainties.
Moreover, a transition of composition (e.g., from light to heavy) will
show up as a characteristic change of the correlation with energy.
Thus, while not providing a measure of the absolute average mass of
cosmic rays, we find the correlation between $X_{\max}$ and $N_{\mu}$ to be a
fairly robust measure of significant, complementary characteristics
of the cosmic-ray composition.

The structure of the paper is as follows. In Sec.~\ref{sec-observables},
we define the observables and the correlation factor.
The analysis in Sec.~\ref{sec-analysis} first regards the case of an
ideal detector, and then the effects of a realistic detector,
of model uncertainties and of different composition mixtures.
Sec.~\ref{sec-discussion} provides a further discussion of the
method. The paper is concluded in Sec.~\ref{sec-conclusion}.


\section{The $X_{\max}$ and $N_{\mu}$ shower observables}
\label{sec-observables}

For our discussion here, we assume that $X_{\max}$ and $N_{\mu}$ are 
measured with different detection systems,
such that the measurement errors are not correlated.
For example, $X_{\max}$ could be 
measured with an array of telescopes that detect the 
fluorescence and/or Cherenkov light created by the air shower,
and $N_{\mu}$ could be measured by an array of muon detectors 
deployed at ground level. 
These techniques have been successfully used to measure showers in the UHE 
range \cite{FlysEye, HiRes, AugerFD, Yakutsk2010, Amiga}.

We assume that both measurement techniques are employed on the same set of showers.
As in Ref.~\cite{XmaxPaper}, we define $X_{\max}$ to be the atmospheric depth where the energy deposit is maximum.
We define $N_\mu$ to be the total number of muons above 1 GeV that reach ground level.
Our results are not strongly affected by
this choice of an energy cut.

It is typical for the muon shower size to be quoted as the number of muons per unit area at a 
certain distance from the shower axis, with the distance being optimized for a particular detector (e.g., \cite{Yakutsk2010, Amiga}). 
Such a muon density observable is highly correlated with $N_\mu$. We use $N_\mu$ because it is 
the more general shower observable (i.e., the specification of a certain shower distance
is not required). Furthermore, an accurate simulation of the lateral development of the shower
is not required. This removes any potential inaccuracies introduced by thinning algorithms \cite{HansenaFluctuations, Knapp:2002vs} (common to
many 3-dimensional air shower simulation codes).

The muon shower size must also be specified at a certain atmospheric depth.
Here, we assume for simplicity that $N_\mu$ is measured at a
constant slant depth of 1200 $\mbox{g}/\mbox{cm}^{2}$,
which is well past the range of $X_{\max}$ observed by
the Auger Observatory \cite{XmaxPaper}.
We checked that the main findings of this study do not depend
on the specific choice of the observation level of $N_\mu$, as long as
$N_\mu$ is observed past shower maximum.
We take the zenith angle of the showers to be a constant $45^{\circ}$,
which corresponds to a slant depth of $\sim1200~\mbox{g}/\mbox{cm}^{2}$ if
the muon detector is located $\sim1400$ m above sea level.
In reality, there would be a distribution of zenith angles
(or equivalently, of slant depths).
In this case, the dependence of $N_\mu$ on zenith angle
should be removed by normalizing each $N_\mu$ measurement
to a central zenith angle
using a parametrized function derived from the complete $N_\mu$ data set.

We define the correlation factor $r$ for a set of showers to be the linear correlation between 
the observed values of $X_{\max}$ and $N_{\mu}$
\begin{equation}
  r = \frac{\mbox{cov}(X_{\max}, N_\mu)} {\sigma(X_{\max}) \sigma(N_\mu)},
  \label{eq:C}
\end{equation}
where cov and $\sigma$ denote the covariance and standard deviation operators, respectively.

\section{Analysis}
\label{sec-analysis}

For our analysis we use the Conex \cite{Conex1, Conex2} air shower simulation package.
Conex uses a state-of-the-art hybrid calculation scheme \cite{Drescher}.
That is, it uses an explicit Monte Carlo simulation for the highest energy interactions in the air shower (i.e., the first several interactions)
and nuclear-electro-magnetic cascade equations for the lower energy interactions.
This allows for a fast, 1-dimensional air shower simulation.
Conex has been
shown to reproduce well the results \cite{Conex1} (e.g., the mean and the natural
shower-to-shower fluctuation of the number of showers particles vs. depth) of the
CORSIKA \cite{Corsika} air shower simulation code.

\subsection{Ideal Detector}

Let us first consider the case where the detectors are ideal, i.e., the detectors 
have zero measurement uncertainty.
In Fig.~\ref{fig:1}a, we show a scatter plot of $X_{\max}$ and $N_{\mu}$  values
for air showers initiated by protons and iron nuclei with
energy of $10^{19}$ eV (1000 events each).
The showers were simulated with Conex using the
QGSJET-01~\cite{QGSJET} high energy interaction model.

With the consideration of an ideal detector, protons and iron nuclei are well separated
on the $N_{\mu}$-$X_{\max}$ plane.
Proton showers produce the broader distribution; iron showers produce the
more narrow distribution.
For the proton distribution, the correlation factor is $r_{p} = 0.0$.
For the iron distribution, $r_{Fe} = 0.7$.
For the union of the two sets, $r_{p+Fe} = -0.51$.

The main result we want to elucidate is the following.
For composition mixtures that contain large fractions
of two or more chemical species that are well separated in mass number
(i.e., significant fractions of both light and heavy nuclei -- a well-mixed composition), 
the value of $r$ will be significantly negative.
This is in contrast to when the composition is comprised of only one
chemical species, for which case the value of $r$ is near zero or positive.

Note that the result that $r$ is negative for well-mixed compositions
depends mainly on the relative separation between protons and heavy nuclei with regard
to the two shower observables.
For the same initial conditions, $X_{\max}$ is
$\sim 100$ g/cm$^2$ deeper for proton showers than for iron showers, and $N_{\mu}$ is
$\sim40\%$ greater for iron showers than for proton showers.
Both of these expectations are not strongly dependent on the details of the high energy hadronic interactions.
In this way, the connection between a negative value of $r$ and the composition mixture is
not strongly model dependent.

As an aside, we note there is also a ``geometrical'' correlation between $X_{\max}$
and $N_{\mu}$: For measurements past the shower maximum at fixed slant depth,
$N_{\mu}$ will increase
for deeper $X_{\max}$ showers. This effect shows up particularly well in the case of iron
and its relatively large value $r_{Fe} = 0.7$. The effect will be diluted,
however, for realistic detector conditions.

\subsection{Realistic detector}

We take our nominal $X_{\max}$ measurement uncertainty to be 20 $\mbox{g}/\mbox{cm}^{2}$.
This is similar to the measurement uncertainty of $X_{\max}$ with the fluorescence detector of the
Pierre Auger Observatory \cite{XmaxPaper}.
We take our nominal $N_{\mu}$ measurement uncertainty to be 20\%.
This is similar to the resolution of the muon density observable $\rho_{\mu}$ used with the Yakutsk \cite{Yakutsk2010}
detector arrays. The AMIGA \cite{Amiga} detector array is expected to have a similar resolution.

\begin{figure*}
\begin{center}
 \epsfig{file=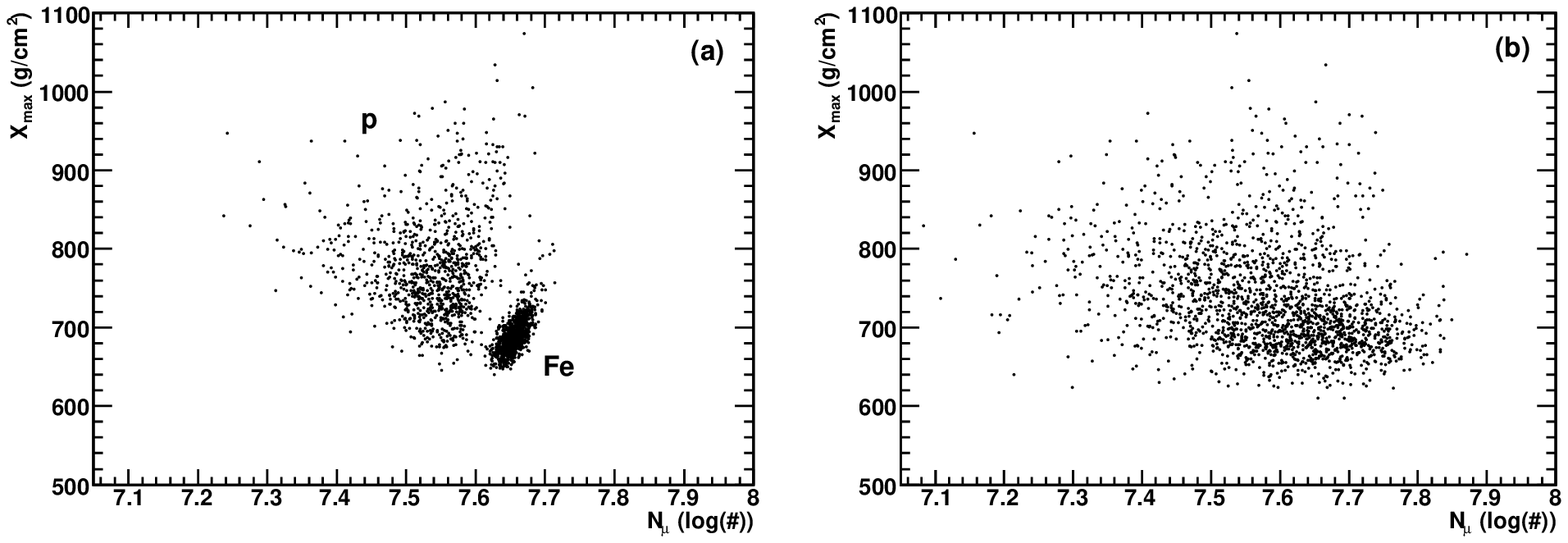, width=1.0\linewidth}
 \caption{$X_{\max}$ - $N_{\mu}$ distribution for
proton showers and iron showers with an
energy of $10^{19}$ eV and a zenith angle of $45^\circ$. There are 1000 iron showers and 1000 proton showers plotted.
The showers were simulated with Conex using the QGSJET-01 high energy interaction model.
Figure (a) is for an ideal detector (i.e., zero measurement uncertainty).
Figure (b) is for a realistic detector; see the text for a description of the measurement uncertainties.}
 \label{fig:1}
  \end{center}
\end{figure*}

In Fig.~\ref{fig:1}b, we show a scatter plot of $X_{\max}$ and $N_{\mu}$ values,
which includes our nominal measurement uncertainties.
As before, the distributions are for air showers initiated by protons and iron nuclei with an
energy of $10^{19}$ eV.

In contrast to the ideal detector consideration, protons and iron nuclei are 
not well separated in Fig.~\ref{fig:1}b.
For the proton distribution, the value of $r$ remains $r_{p} = 0.0$.
For the iron distribution, the value of $r$ decreases, $r_{Fe} = 0.1$.
For the union of the two sets, the value of $r$ increases, $r_{p+Fe} = -0.32$.

Because $X_{\max}$ and $N_{\mu}$ are measured with independent detectors,
the effect of measurement uncertainty is to de-correlate the observables.
That is, the values of $r$ are closer to zero than in the
case of the ideal detector.
However, the main observation from before remains valid:
the value of $r$ can discriminate between a well-mixed composition
and a pure composition.

The discrimination power of $r$ holds also for data sets smaller
than the one shown in Fig. 1.
For example, for a data set of 200 events,
the 1-sigma statistical uncertainty on $r$ is $\sim0.07$.
For this case, $r_{p}$ and $r_{p+Fe}$ are separated by nearly 5-sigma.

\subsection{Effects of measurement resolution and model uncertainties}

\begin{figure*}
\begin{center}
 \epsfig{file=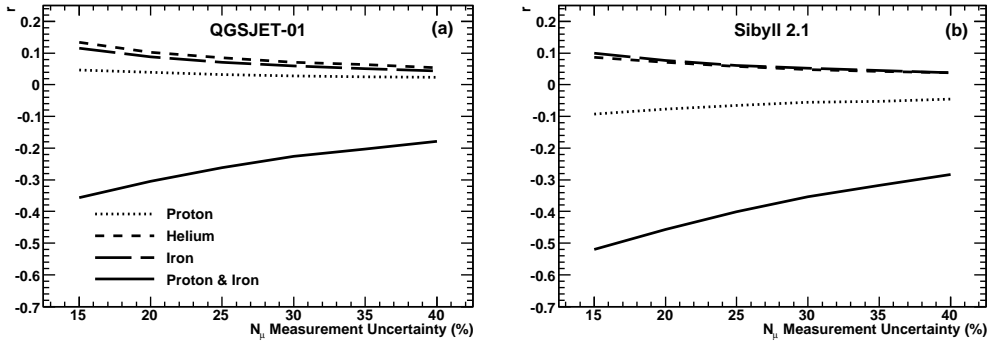, width=1.0\linewidth}
 \caption{Correlation factor $r$ versus the $N_{\mu}$ measurement uncertainty
for three different pure compositions and one well-mixed composition, a proton-iron mixture.
The proton-iron mixture is 50\% protons and 50\% iron nuclei. 
The results in Figure (a) are derived with the QGSJET-01 interaction model.
The results in Figure (b) are derived with the Sibyll 2.1 interaction model.}
 \label{fig:2}
  \end{center}
\end{figure*}

The ability of $r$ to discriminate between composition mixtures 
is dependent on the measurement uncertainties.
In Fig.~\ref{fig:2}, we show the value of $r$ for
different composition scenarios
versus the measurement uncertainty of the $N_{\mu}$ observable,
keeping the $X_{\max}$ uncertainty at 20 $\mbox{g}/\mbox{cm}^{2}$. 
The results are for $10^{19}$ eV showers.
Fig. 2a is calculated with the QGSJET-01 interaction model, while
Fig. 2b is calculated with the Sibyll 2.1 interaction model.
Out of the four models considered in this paper -- QGSJET-01, QGSJET II-03 \cite{QGSJETII}, Sibyll 2.1, and EPOS --
QGSJET-01 and Sibyll 2.1 give the minimum and maximum values of $r_{p+Fe}$,
respectively.

Fig.~2 demonstrates that the separation between a pure composition and a well-mixed
composition is significant over a broad range of $N_{\mu}$
measurement uncertainties.

Fig.~2 also demonstrates how the value of $r$ depends on the interaction model used 
in the simulation. Comparing QGSJET-01 and Sibyll 2.1, the values of $r$
are similar for both helium and iron nuclei. 
Indeed, we have found that for all pure beams with $A \geq 4$, $r$ is similar for
all four interaction models, when realistic measurement uncertainties are taken
into account.

However, the situation for protons is different.
The value of $r_{p}$ is somewhat model dependent.
Related to this, the value of $r$ for mixtures
that contain a significant proton fraction is also model dependent.
The value of $r_{p}$ calculated with either the Sibyll 2.1 or EPOS models
is significantly more negative than the value of
$r_{p}$ calculated with either of the QGSJET models.
Without going into the details of the hadronic interactions themselves,
we offer the following observation as an insight into the cause of this difference.
For the Sibyll 2.1 and EPOS models, relative to the QGSJET models,
it is more common for a large faction of the primary's energy to be
transferred to electromagnetic particles early on in the development
of the shower. 
This feature tends to produce proton showers that are
somewhat photon-like (i.e., a deep $X_{\max}$ and small $N_{\mu}$).
In turn, the presence of these photon-like showers tends to make
the value of $r$ more negative for protons.

As also discussed below,
the model uncertainty reduces the discrimination power of $r$,
but useful constraints are still possible.
For instance, with our nominal measurement uncertainties and
for a data set of 200 events,
an observation of $r < -0.25$ will favor a well-mixed composition
independent of interaction models.


\subsection{Sensitivity to different composition mixtures}

\begin{figure*}
\begin{center}
 \epsfig{file=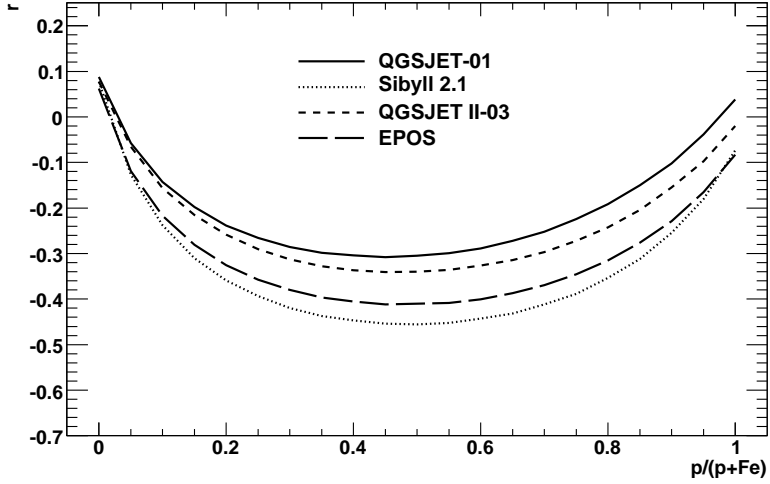, width=0.8\linewidth}
 \caption{Correlation factor $r$ for a proton-iron composition 
mixture. The quantity $p/(p+Fe)$ is the ratio of the number of protons in the beam to the total number of particles. 
We show the results of four different interaction models.
We assume a realistic detector with the nominal measurement uncertainties described in the text.}
 \label{fig:3}
  \end{center}
\end{figure*}

In Fig.~\ref{fig:3}, we show the value of $r$ for a proton-iron mixture as a
function of the number ratio $p/(p+Fe)$.
The value of $r$ is calculated with our nominal measurement uncertainties.
We show results for all four interaction models.

The minimum value of $r$ occurs for a ratio $p/(p+Fe) \approx 0.5$.
Near this ratio the value of $r$ is fairly flat.
For example, in the range $0.2 < p/(p+Fe) < 0.8$, $r$ changes by $<0.1$.
This is true for all models.
For well-mixed proton-iron mixtures, the value of $r$ 
can differ by up to $\sim0.18$ between the models.
At higher (lower) proton fractions, $r$ increases sharply
and approaches $r_{p}$ ($r_{Fe}$).
A value of $r$ near zero is indicative of a fairly pure composition,
while a more negative value of $r$ occurs only for a well-mixed beam.

\begin{figure*}
\begin{center}
 \epsfig{file=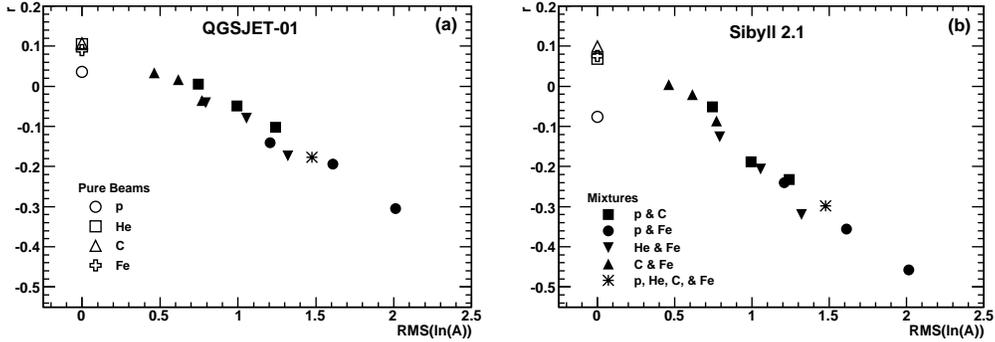, width=1.0\linewidth}
 \caption{Correlation factor $r$ for several different composition
mixtures versus RMS(ln(A)). For each bi-species beam, we plot three different mixture ratios:
80\%-20\%, 50\%-50\%, and 10\%-90\%, where the first percentage is the fraction
of the light component. For the quad-species beam, we test the case where all four species are 
present in equal measure.
We assumed a realistic detector with the nominal measurement uncertainties described in the text.}
 \label{fig:4}
  \end{center}
\end{figure*}

In Fig.~\ref{fig:4}, we show the correlation factor for several different
composition mixtures calculated with our nominal measurement uncertainties.
In Fig.~\ref{fig:4}a, the calculations were performed with the
QGSJET-01 interaction model, while in Fig.~\ref{fig:4}b
the calculations were performed with the Sibyll 2.1 interaction model.
In most cases, these two models
bracket the values of $r$ calculated with the QGSJET II-03 and EPOS models.
We plot $r$ as a function of $\mbox{RMS}(\mbox{ln}(A)) = \sqrt{\mbox{var}(\mbox{ln}(A))}$,
where $A$ is the mass number of the cosmic rays present in the beam
and var is the variance operator.
The quantity RMS(ln($A$)) is a measure of the purity of the beam
(i.e., a pure beam will have RMS(ln($A$)) = 0, and a mixed beam
will have $\mbox{RMS}(\mbox{ln}(A)) > 0$).

We show four different pure beams: p, He, C, and Fe;
four different bi-species beams: p \& C, p \& Fe, He \& Fe, and C \& Fe;
and one quad-species beam: p \& He \& C \& Fe.
For each bi-species beam, we plot three different mixture ratios:
80\%-20\%, 50\%-50\%, and 10\%-90\%, where the first percentage is the fraction
of the light component.
For the quad-species beam, we test the case where all four species are 
present in equal measure.

Over a wide variety of mixtures, the correlation between $r$ and
RMS(ln($A$)) is
quite strong: $r$ decreases as RMS(ln($A$)) increases.
The most negative value of $r$ occurs when the mixture is
dominated by nearly equal portions of proton and iron nuclei.


From an inspection of Fig.~\ref{fig:4},
we see that an observation of $r < -0.25$ would
indicate a well-mixed beam, i.e., $\mbox{RMS}(\mbox{ln}(A)) \gtrsim 1.3$.
In turn, an observation of $r > -0.05$ would
indicate a fairly pure beam, i.e., $\mbox{RMS}(\mbox{ln}(A)) \lesssim 0.7$.
This indication is independent of the hadronic model.  

In contrast, such a model independent indication of the composition seems impossible to achieve from the $N_\mu$ observable alone (i.e.,
the model uncertainties dominate everything). An example of this was given in the introduction.
Let us give one more example here.
We show in Fig. 5 the RMS spread in $N_\mu$ for the same composition mixtures shown in Fig. 4. 
While there might be a moderate correlation between
RMS($N_\mu$) and RMS(ln($A$)) for a given model, the difference in RMS($N_\mu$) between
models clearly dominates.

Given a 2-d distribution of $N_{\mu}$ and $X_{\max}$ measurements,
the lowest order moment involving $N_{\mu}$ that has a clear interpretation regarding the composition
seems to be the normalized covariance (i.e., $r$). 
There are, of course, other ways to look at this problem. 
One could form a linear combination like $a X_{\max} + b N_{\mu}$. The second moment of this new variable contains $\mbox{cov}(X_{\max}, N_\mu)$, 
and, if formulated carefully, one should expect this new variable to have some similarities to $r$ with regard to composition analysis.
Notwithstanding the full range of more complex formalizations, we have found the $r$ variable both to be simple and to have a clear, meaningful interpretation.


\section{Further discussions}
\label{sec-discussion}

We have investigated the use of the Spearman rank correlation coefficient as a substitute for $r$. 
We found no significant improvement.

In Section 3, we only considered showers with an energy of $10^{19}$ eV.
We checked that the dependence of $r$ on the composition mixture is similar
throughout the UHE regime.

When calculating $r$ from data (i.e., not simulations), the showers will be distributed within an energy bin.
In this case,
the $X_{\max}$ and $N_{\mu}$ observables must be normalized to a central energy.
If the energy estimate for the shower has a statistical uncertainty of less than 10\%
(expected if the longitudinal profile of the shower is well measured \cite{ESpecPaper}), 
this normalization procedure does not add appreciably to the measurement uncertainty of
either $X_{\max}$ or $N_{\mu}$.

As an observable,
the value of $r$ is rather robust against systematic errors associated with
the detectors. For example, a constant offset or multiplicative factor 
in the measurement of $X_{\max}$ or
$N_\mu$ does not affect the calculation of $r$.

If there is a light to heavy composition transition, as indicated by
recent elongation rate measurements, then there will likely be an energy
range where the composition is well-mixed.
The work presented here indicates that it is possible to obtain
a robust, independent indication
of a well-mixed condition.
The discrimination power of $r$ is best between a
pure composition and a composition with significant fractions of both protons and iron nuclei.
It is interesting that such a well-mixed scenario, near $10^{19}$ eV, has been suggested in
the literature \cite{Allard2008}. 

It is possible to measure $r$ as a function of energy (i.e., in consecutive
energy bins).
If there exists good statistics over perhaps an order of magnitude in energy,
it may be possible to observe the composition entering and/or exiting a well-mixed state,
i.e., to observe $r$ changing with energy.
As an example, consider the case where the composition is transitioning from a nearly pure
beam of protons (or iron) to a well-mixed beam of protons and iron. In this case, 
the expected change of $r$ with energy can be obtained from Fig.~\ref{fig:3}.
The observation of such a change in $r$ with energy would be a particularly strong, 
model independent indication of a composition transition.

For a given interaction model to provide a successful description of the shower data,
there must be consistency between the
range of RMS(ln($A$)) implied by the observed $\left< X_{\max} \right>$ and RMS($X_{\max}$)
and the range of RMS(ln($A$)) implied by the observed $r$.
For example, suppose that $\left< X_{\max} \right>$ is observed to be
somewhat less than $\left< X_{\max} \right>_{proton}$ and that RMS($X_{\max}$)
is observed to be broad such
that the only composition mixtures that fit well 
are well-mixed.
Then if the observed value of $r$ is inconsistent with a well-mixed composition,
the interaction model cannot be judged to be self-consistent.
In this way, a measurement of $r$ can be used
to evaluate the self-consistency of the interaction models.

An observed value of $r$ inconsistent with a pure composition,
in the framework of the models discussed here,
does not exclude the possibility that 
the composition is actually quite pure
and that there are
physical processes that are currently not accounted for in air shower simulations.
However, the observed value of $r$ would put constraints on any hypothesized physical processes
that are proposed to explain, e.g., the observed elongation rate.



\section{Conclusions}
\label{sec-conclusion}

We have studied in detail the sensitivity of the correlation factor $r$
between $X_{\max}$ and $N_{\mu}$ to qualities of the
UHE cosmic ray composition.
%
The correlation factor provides composition information that complements
the standard composition analysis based on the average and RMS of the
observables.
It incorporates the muon shower size observable $N_{\mu}$ and thus
provides an indication of the composition that is experimentally independent
of the elongation rate analysis.
While not providing a direct indication of the average nuclear mass,
the value of $r$ is sensitive to whether the UHE cosmic ray beam is
pure or well-mixed.
Testing whether the beam is pure or well-mixed is 
well motivated by recent measurements of the depth of shower maximum.
The discrimination power of $r$ is affected by model uncertainties; however,
it is possible to make useful constraints independent of the interaction models.
More sophisticated analysis methods might be developed in the future
that make a combined use of the information contained in the
moments of the two-dimensional $X_{\max}$ - $N_{\mu}$ - distribution,
particularly if the detector effects of a specific experimental
setup are well understood and if the theoretical model uncertainties
can be reduced. Here we showed what kind of new information is contained
in the correlation parameter $r$.

\section{Acknowledgements}
We wish to thank our fellow Pierre Auger collaborators and the members of our research group at the
University of Siegen.
The work has been supported by the German Federal Ministry of
Education and Research (BMBF 05A08PS1).
P.Y. gratefully acknowledges the support of the Alexander von Humboldt Foundation,
the National Science Foundation (award number PHY-0838088), 
the U.S. Department of Energy through the LANL/LDRD Program,
and the Michigan Space Grant Consortium.


\newpage

\begin{figure*}
\begin{center}
 \epsfig{file=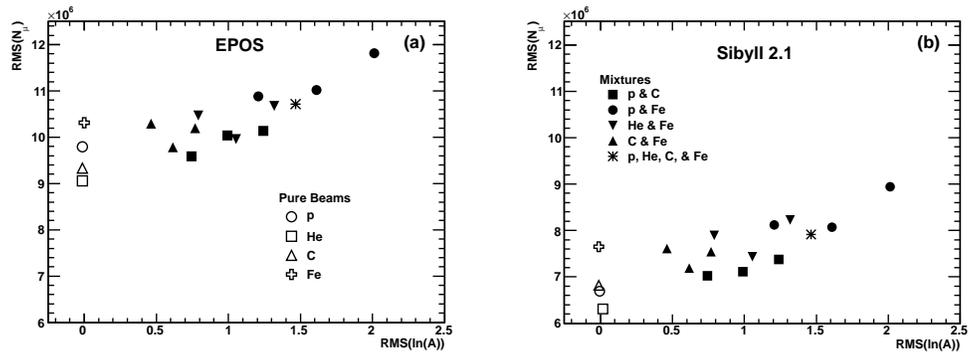, width=1.0\linewidth}
 \caption{RMS($N_\mu$) for several different composition
mixtures versus RMS(ln(A)).
We assumed a realistic detector with the nominal measurement uncertainties described in the text.}
 \label{fig:5}
  \end{center}
\end{figure*}

\end{document}